\documentstyle[editedvolume,numreferences]{crckapb}

\newcommand{\be}{\begin{equation}}
\newcommand{\ee}{\end{equation}}
\newcommand{\bea}{\begin{eqnarray} \nonumber}
\newcommand{\eea}{\end{eqnarray}}
\newcommand{\beaa}{\begin{eqnarray}}
\newcommand{\eeaa}{\end{eqnarray}}
\newcommand{\ba}{\begin{array}}
\newcommand{\ea}{\end{array}}
\newcommand{\bit}{\begin{itemize}}
\newcommand{\eit}{\end{itemize}}
\newcommand{\ben}{\begin{enumerate}}
\newcommand{\een}{\end{enumerate}}
\newcommand{\bib}{\bibitem}
\def\bib{\bibitem}

\def\lan{\langle}

\def\le{\left}

\def\Lrar{\Leftrightarrow}

\def\pa{\partial}
\def\ran{\rangle}
\def\rar{\rightarrow}

\def\ri{\right}
\def\ti{\tilde}

\def\al{\alpha}

\def\de{\delta}
\def\De{\Delta}
\def\ep{\epsilon}

\def\te{\theta}

\def\la{\lambda}
\def\La{\Lambda}
\def\si{\sigma}

\def\vec#1{{\bf #1}}

\newcommand{\mlab}[1]{\label{#1}}

\begin{opening}
\title{DEFECT FORMATION THROUGH BOSON CONDENSATION\protect\\
IN QUANTUM FIELD THEORY}

\author{GIUSEPPE VITIELLO}
\institute{Dipartimento di Fisica and INFM Unita' di Salerno \\ 
Universit\`a di Salerno, 84100 Salerno, Italy\\
vitiello@vico.phys.unisa.it}

\end{opening}

\runningtitle{DEFECT FORMATION THROUGH BOSON CONDENSATION}

\begin{document}


\section{Introduction}

The study of many body physics as well as the study of elementary    
particle physics has convinced us that at a very basic level    
Nature is ruled by quantum dynamical laws. On the other hand, we    
also know and observe several systems, such as    
superconductors, superfluids, crystals, ferromagnets,    
which behave as {\it macroscopic} quantum systems.   
   
The question then arises of how the quantum dynamics may    
generate the observed macroscopic properties. In other words,    
how it happens that the macroscopic scale characterizing those    
systems is dynamically generated out of the microscopic    
scale of the quantum elementary components\cite{Um2}.    
   
Moreover, we also observe a varieties of phenomena where quantum    
objects coexist and interact with extended macroscopic objects     
which show classical behavior, e.g. vortices in    
superconductors and superfluids, magnetic domains in    
ferromagnets, dislocations and other {\it defects}  
in crystals. 

Thus, we are faced also with the question of the quantum origin    
of the macroscopically behaving extended objects and of their    
interaction with quanta\cite{Um1}. Even for structures at cosmological    
scale, the question of their dynamical origin from elementary    
components asks for an answer consistent with  quantum    
dynamical laws\cite{kib}.   
   
Macroscopic quantum systems are quantum systems not, of course,   
in the rather trivial sense that they are made by quantum components,    
but in the sense that, although they behave classically, 
nevertheless some of    
their macroscopic features cannot be understood without recurse    
to quantum theory. Quantum theory thus appears not confined to    
microscopic phenomena.   
   
In this respect it is remarkable that these "classical" systems    
present observable ordered patterns, e.g. crystal ordering,    
phase coherence, ferromagnetic ordering, etc.. Moreover, most    
extended objects present some topological singularity, and    
interesting enough, these topologically non-trivial defects    
only appear in systems presenting an ordered state.   
   
The formation of defects in the course of phase    
transitions provides a further source of questions which are    
attracting much attention since it appears that defect   
formation during phase transitions may reveal unifying    
understanding of phenomena belonging to a wide range of energy    
scale\cite{zu}.   
   
The task of this paper is to review some of    
the main aspects in the Quantum Field Theory (QFT)    
description of topological defect formation, which also illustrate how to
get the macroscopic scale out of the quantum dynamics.  I will further    
mention some recent developments dealing with temperature effects    
on defect formation\cite{ro}.   
   
The paper is organized as follows. I will consider    
the problem of dynamical generation of order in quantum systems 
in Section 2.    
The key ingredients are the mechanism of spontaneous breakdown of    
symmetry (SBS) and the consequent appearance of Nambu-Golsdtone    
(NG) boson particles\cite{go,nj} (such as phonons in crystals). 
In order to    
present a general, model independent discussion, I will    
use functional integration techniques. As we will see, NG modes    
manifest as long range correlations and thus they are responsible    
of the above mentioned change of scale, from microscopic to    
macroscopic. The coherent boson condensation of    
NG modes turns out to be the mechanism by which order is    
generated. From the point of view of the invariance    
properties of the theory, the mathematical    
structure of the contraction\cite{iw} of the symmetry group is the one    
controlling the SBS mechanism\cite{dv}.   
   
I will show how topologically non-trivial defects    
are generated in quantum systems by non-homogeneous boson    
condensation in Section 3. 
Here the so called boson transformation reveals to    
be the crucial tool. I will prove that topological defects only    
can be formed in the presence of NG modes, i.e. in the presence    
of ordering. This sheds some light on the mechanism by which    
defect formation occurs in phase transitions, i.e. in the    
presence of gradients of the order parameter. Interaction of    
defects with quanta is also very briefly considered in this Section.   
Explicit vortex solutions in terms of boson condensation are presented in
Section 4.
  
Temperature effects and volume effects on SBS,    
on defects formation and on symmetry restoration are considered in
Section 5. Contact with the
problem of defect formation in phase transition processes is also made
in this Section.
 
A preliminary remark to my subsequent discussion is the following.    
   
The von Neumann theorem in Quantum Mechanics (QM)
\cite{vn} states that for systems with a finite number of degrees of    
freedom all the representations of the canonical commutation    
relations are unitarily equivalent. This theorem actually states    
that in QM the physical system can only live in one single phase:    
unitary equivalence means indeed physical equivalence and thus    
there is no room ( no representations) to represent different    
physical phases. Fortunately, such a situation drastically    
changes in QFT where systems with infinitely many degrees of    
freedom are studied. In such a case the von Neumann theorem does    
not hold and infinitely many unitarily inequivalent    
representations of the canonical commutation relations do in fact    
exist\cite{brat}. 
It is such a richness of QFT which allows the description    
of different physical phases. The occurrence of spontaneous    
breakdown of symmetry and of the related NG boson condensation    
becomes thus possible in QFT.    
   
Although one can set up many formal devices based on more or    
less sophisticate approximations, or even on semi-classical methods,     
which may nevertheless lead to phenomenologically successful    
results, it should be always understood that the proper    
theoretical framework where to operate dealing with phase    
transitions, defect formation and all that is the larger manifold    
of unitarily inequivalent representations provided by QFT.

\section{Spontaneous breakdown of symmetry and group    
contraction}   

In QFT the dynamics is described by a set of field equations   
for the interacting operator fields, say $\psi(x)$, also called the    
Heisenberg fields. These are the basic fields of the theory    
satisfying equal-time canonical commutation relations and   
the Heisenberg field equations   
\be   
\mlab{bt1}   
\La(\pa) \, \psi(x)\, =\, j[\psi](x)  ~, 
\ee   
where $x\equiv (t,\vec{x})$. $j$ is a functional of the   
$\psi$ field describing the interaction.    
   
Observable phenomena are on the other hand described by    
observable (physical) operator fields (such as phonons),   
say $\varphi (x)$.  They also obey equal-time canonical commutation    
relations and satisfy free field equations, which, with    
convenient care in the renormalization procedure, may be written as
\be   
\mlab{bt2}   
\La(\pa) \, \varphi(x)\, = \, 0~.   
\ee   
The Hilbert space, say ${\cal H}$, for the physical states is    
the Fock space for    
the fields $\varphi$. Solving the dynamical problem thus means to    
compute by means of eq. (\ref{bt1}) the matrix elements of $\psi$ in the    
space ${\cal H}$: this will relate the basic dynamics to the    
observable properties of the physical states. In this way we    
obtain the {\it dynamical map} between interacting fields and physical    
fields\cite{Um2,Um1}:   
\be   
\mlab{bt4}   
 \psi(x)\, =\, F \le[x; \varphi \ri]~.   
\ee   
Eq.(\ref{bt4}) is also called the Haag expansion in the LSZ    
formalism\cite{itz}. I have to stress that the equality in 
(\ref{bt4}) is a "weak"    
equality: it must be understood as an equality among    
matrix elements computed in ${\cal H}$.    
   
This a crucial point and a couple of remarks need to be made.   
First, I observe that the set of $\varphi$ fields must be an    
irreducible set; however, it may happen that not all the elements    
of the set are known since the beginning.   
For example there might be composite (bound states) fields    
or even elementary    
quanta whose existence is ignored in a first recognition.    
Then the computation of the matrix elements in physical states    
will lead to the detection of unexpected poles in the Green's functions,    
which signal the existence of the ignored quanta. One thus    
introduces the fields corresponding to these quanta and repeats   
the computation. This way of    
proceeding is called the self-consistent method\cite{Um1}. In this    
connection, I note that   
it is not necessary to have a one-to-one correspondence   
between the sets $\{\psi_i\}$ and $\{\varphi_i\}$. This    
happens in fact when the set $\{\varphi_i\}$ includes composite    
particles.   
   
Another remark is that, as already mentioned, in QFT the Fock space    
for the physical states is not unique: one may have indeed    
several physical phases, e.g. for a metal the normal phase and    
the superconducting phase, and so on. Fock spaces describing    
different phases are unitarily inequivalent spaces and    
correspondingly we have different expectation values for certain    
observables and even different irreducible sets of physical    
quanta; for example, in ferromagnets this set    
includes magnon fields which do not exist in the non-magnetic    
phase, etc.. Thus, finding the dynamical map involves the    
"choice" of the Fock space where the dynamics has to be realized:    
in other words the same dynamics (i.e. same Heisenberg fields and    
same Heisenberg field equations) may generate different physical    
phases.   
   
Suppose now that the dynamics is invariant under some group $G$ of    
transformations of $\psi$:   
\be\mlab{rs1}   
\psi(x) \rar \psi'(x) = g\le[ \psi(x) \ri]~,   
\ee   
with $g \in G$. Invariance of the dynamics means that    
the Heisenberg equations (or the Lagrangian from which they may    
be derived) are invariant (in form) under the   
transformations of $G$.    
   
One says that symmetry is spontaneously broken    
when the vacuum state in the Fock space ${\cal H}$ is not    
invariant under the group $G$ but only under one of its    
subgroups\cite{Um2,Um1,itz}.    
   
Eq. (\ref{bt4}) implies that when $\psi$ is    
transformed as in (\ref{rs1}), then   
\be\mlab{rs3}   
\varphi(x) \rar \varphi'(x) = g'\le[ \varphi(x) \ri]~,   
\ee   
such that    
\be\mlab{rs4}   
g\le[ \psi(x)\ri] = F\le[ g'\le[\varphi(x)\ri] \ri]~,   
\ee   
with $g'$ belonging to some group of transformations $G'$.    
   
Now it happens, as we will see below, that $G \neq G'$ when    
symmetry is spontaneously broken, with $G'$ the group contraction    
of $G$\cite{dv}; when symmetry is not broken $G = G'$. 

Since $G$ is the invariance group of    
the dynamics, eq. (\ref{bt4}) requires that $G'$ is the    
group under which free fields equations are invariant, i.e.    
$\varphi'$ also is a solution of (\ref{bt2}). Since eq.    
(\ref{bt4}) is a weak equality, $G'$ depends on the choice of the    
Fock space ${\cal H}$ (among the physically realizable unitarily    
inequivalent state spaces). Thus we see that the original (same)    
invariance of the dynamics may manifest itself in different    
symmetry groups for the $\varphi$ fields according to different    
choices of the physical state space. Since this    
process is constrained by the dynamical equations (\ref{bt1}), it    
is called the {\it dynamical    
rearrangement of symmetry}\cite{Um2,Um1,vit}.    
   
To be specific, let me    
consider, in the path-integral formalism, 
a complex scalar field $\phi(x)$   
interacting with a gauge field $A_\mu(x)$ (Anderson-Higgs-Kibble type 
model)\cite{and,hig,ki}. The
lagrangian density ${\cal    
L}[\phi(x), \phi^*(x), A_\mu(x)]$ is invariant under the global    
and the local gauge transformations:   
\be\mlab{lp1}   
\phi(x) \rar e^{i \te} \phi(x) \qquad, \qquad \qquad   
A_\mu(x) \rar A_\mu(x)~,   
\ee   
\be\mlab{lp2}   
\phi(x) \rar e^{i e_0 \la(x)} \phi(x) \qquad, \qquad \qquad   
A_\mu(x) \rar A_\mu(x) \, + \, \pa_\mu \la(x)~,   
\ee   
respectively, where $\la(x)\rar 0$ for $|x_0|\rar \infty$ and/or    
$|{\bf x}|\rar \infty$. The Lorentz gauge 
$\pa^\mu A_{\mu H}(x)\,=\,0$ is used. I put   
$\phi(x)=\frac{1}{\sqrt{2}}\le[\psi(x) + i \chi(x)\ri]$.

Spontaneous breakdown of symmetry is introduced through the condition      
$\lan 0| \phi_H(x)|0\ran \equiv {\ti v} \neq 0$,   
with ${\ti v}$ constant and I put
$\rho(x) \equiv \psi(x) - {\ti v}$. Here $\phi_H(x)$ denotes the
Heisenberg field. $\phi(x)$ is the c-number field entering 
the functional integral.
The generating functional, including the gauge   
constraint, is\cite{MPUV75}   
\bea   
&&W[J,K]=\frac{1}{N}\int [d A_\mu] [d\phi] [d\phi^*] [d B]\,    
\exp\le[i\int d^{4} x\le({\cal L}(x) + B(x) \pa^\mu A_\mu(x) + \ri.\ri.   
\\ \mlab{lp5}   
&&\qquad \qquad \qquad    
\le.\le. K^* \phi + K \phi^* + J^\mu(x) A_\mu(x)+ i {\ep}   
|\phi(x) -v|^2 \ri)\ri]  ~,  
\eea   
with $N$ a convenient normalization. $B(x)$ is an auxiliary field   
which guarantees the gauge condition.
The r\^ole of the $\ep-$term    
is to specify the condition of    
symmetry breakdown under which we want to compute the    
path-integral\cite{MPU741,MPU742}. It may be given the physical 
meaning of the small    
external field triggering the symmetry breakdown. The limit    
$\ep \rar 0$ must be made at the end of the computations.    

As customary, I will use the notation   
$\lan F[\phi]\ran_{K,\ep}$ 
to denote functional average and   
$\lan F[\phi]\ran_{\ep} \equiv  \lan F[\phi]\ran_{\ep, K=0}~,~   
\lan F[\phi]\ran \equiv  \lim_{\ep \rar 0} \, \lan F[\phi]\ran_{\ep}$.  
Note that $\lan\chi(x)\ran_\ep=0$ because of the invariance under    
$\chi\rar - \chi$. 

Invariance of the path-integral under
the change of variables   
(\ref{lp1}) (and/or (\ref{lp2}) ) leads to
\be\mlab{gp6}   
\lan \psi(x)\ran_\ep =   
\sqrt{2}\,\ep v \int d^4y \lan \chi(x)\chi(y)\ran_{\ep} = \sqrt{2}\,\ep   
v\, \De_\chi(\ep,0)~.   
\ee   
This is one of 
the Ward-Takahashi identities. Such identities    
carry the symmetry content of the theory.
In momentum space   
the propagator for the Heisenberg field $\chi$ has the general form   
\be\mlab{gp8}   
\De_\chi(0,p) = \lim_{\ep \rar 0}\le[ \frac{Z_\chi}{p^2-m_\chi^2+i\ep   
a_\chi} + (continuum \;contributions) \ri]\, .   
\ee   
$Z_\chi$ and $a_\chi$ are renormalization constants.   
The integration in eq.(\ref{gp6}) picks up the pole contribution at   
$p^2=0$, and leads to   
\be\mlab{gp9}  
{\ti v}= \sqrt{2}\frac{Z_\chi}{a_\chi} v \Lrar m_\chi = 0   ~,
~~~~or ~~~~
{\ti v}= 0 \Lrar m_\chi \neq 0   ~.
\ee   
The Goldstone theorem\cite{go} is thus proved\cite{MPU741,MPU742}: 
if the symmetry is   
spontaneously broken (${\ti v} \neq 0$),    
a massless mode exists, whose interpolating Heisenberg field    
is $\chi(x)$. It is the NG boson mode. Since it is massless it    
manifests as a long range correlation mode. Notice that 
the NG mode is an    
elementary field. In other models it may appear as a bound state,   
e.g. the magnon in ferromagnets\cite{sha}.   
Note that\cite{MPU741,MPU742}  
 ${\ti v}$ is independent of $|v|$,   
although the phase of  $|v|$ determines the one of ${\ti v}$: 
as in ferromagnets, once an    
external magnetic field is   
switched on, the system is magnetized independently of the    
strength of the external field.   

The analysis of the two-point functions of the theory shows\cite{MPUV75}
that the   
model contains a massless negative norm state (ghost), besides the    
NG massless mode $\chi$, and 
a massive vector field $U^{\mu}$.   
The dynamical maps are:   
\be\mlab{lp56a}   
S\,=\,: S[\rho_{in}, U^\mu_{in}, \pa(\chi_{in} - b_{in})]:  ~,  
\ee   
\be\mlab{lp57a}   
\phi_H(x)= :\exp\le\{i  \frac{Z_\chi^{\frac{1}{2}}}   
{ {\ti v}}\chi_{in}(x) \ri\}     
\le[{\ti v} + Z_\rho^{\frac{1}{2}} \rho_{in}(x) + F[\rho_{in},   
U^\mu_{in}, \pa(\chi_{in} - b_{in})] \ri]:~,
\ee
\be\mlab{lp57b}   
A^{\mu}_{H}(x)= {Z_3^{\frac{1}{2}}} U^{\mu}_{in}(x) +   
{\frac{Z_\chi^{\frac{1}{2}}}{e_0{\ti v}}} \pa^\mu b_{in}(x)+ 
:F^{\mu}[\rho_{in},   
U^\mu_{in}, \pa(\chi_{in} - b_{in})] :~, 
\ee   
where the functionals $F$ and $F^\mu$ are to be determined within a   
particular model. 
These relations are weak equalities and are equivalent to the 
familiar LSZ reduction formula\cite{itz}. 
It will be also used    
$A^{0\mu}_{H}(x) \equiv A^{\mu}_{H}(x) -  
{ e_0{\ti v}}:\pa^\mu b_{in}(x):$.  
In eqs. (\ref{lp56a})-(\ref{lp57b})    
$\chi_{in}$ denotes the NG mode, $b_{in}$ the ghost mode,    
$U^{\mu}_{in}$ the massive vector field and $\rho_{in}$ the massive    
matter field. Their field equations are   
\be\mlab{lp24}   
\pa^2 \chi_{in}(x)\,=\,0~,~~~   
\pa^2 b_{in}(x)\,=\,0~, ~~~  
(\pa^2 \, + \, m_\rho^2)\rho_{in}(x) \, =\,0~,   
\ee   
\be\mlab{lp29}   
( \pa^2 \, + \, {m_V}^{2}) U_{\mu in}(x) \, =\, 0 \qquad, \qquad   
\pa^\mu U_{\mu in}(x) \, =\, 0~.   
\ee   
with ${m_V}^{2} =   
\frac{Z_{3}}{Z_\chi}   
(e_0{\ti v})^{2}$. We also have   
\be\mlab{lp43}   
B_H(x)= \frac{e_0 {\ti v}}{Z_\chi^{\frac{1}{2}}}[b_{in}(x) -   
\chi_{in}(x)]  \, ~.
\ee   
The field equations 
for $B_{H}$ and $A_{H \mu}$ are   
\be\mlab{lp37}   
\pa^2 B_H(x)\, =\,0   ~,~~~
- \pa^2 A_{H \mu}(x) \, =\, j_{H \mu}(x) \, -\,    
\pa_\mu B_H(x) ~,  
\ee   
with $j_{H \mu}(x)= \de{\cal L}(x)/\de A^{\mu}_H(x)$.   
One may then require that the current $j_{H \mu}$ is the only source of   
the gauge field $A_{H \mu}$ in any observable process. This amounts to   
impose the condition: $_p\lan b|\pa_\mu B_H(x)|a\ran_p\,   
= \,0$, i.e.  
\be\mlab{lp45}   
(- \pa^2) \,_p\lan b| {A^{0}}_{H \mu}(x) |a \ran_p \, =   
\,_p\lan b| j_{H   
\mu}(x) |a\ran_p  ~,
\ee   
where $|a\ran_p $ and   $|b\ran_p $ denotes two generic physical   
states. Eq.(\ref{lp45}) are the classical Maxwell equations. The condition
$_p\lan b|\pa_\mu B_H(x)|a\ran_p\,   
= \,0$ leads to the Gupta-Bleuler-like condition   
\be\mlab{lp49}   
[\chi_{in}^{(-)}(x)  \, - \,  b_{in}^{(-)}(x)]|a\ran_p\, = \,0   ~, 
\ee   
where $\chi_{in}^{(-)}$ and $b_{in}^{(-)}$ are the positive-frequency   
parts of the corresponding fields. Thus we see that $\chi_{in}$ and   
$b_{in}$ do not participate to any observable reaction. Note in    
fact that     
they   
are present in the $S$ matrix in the combination $(\chi_{in} -   
b_{in})$. It is to be remarked that    
the NG boson do not disappear from the theory: we will see that   
in situations in which the vacuum is not translationally invariant, the   
NG fields can have observable effects.   
   
The study of the dynamical rearrangement of the symmetry shows that 
local gauge transformations of the   
Heisenberg fields   
\be\mlab{lp50}   
\phi_H(x) \rar e^{i e_0 \la(x)} \phi_H(x)~,~~ ~  
A_H^\mu(x) \rar A_H^\mu(x) \, + \, \pa^\mu \la(x)~,~~ ~  
 B_H(x) \rar B_H(x)   
\ee   
are induced by the in-field transformations   
\be\mlab{lp51a}   
\chi_{in}(x)  \rar   \chi_{in}(x) \, + \, \frac{e_0 {\ti   
v}}{Z_\chi^{\frac{1}{2}}} \la(x) ~,~~ ~  
b_{in}(x)  \rar   b_{in}(x) \, + \, \frac{e_0 {\ti   
v}}{Z_\chi^{\frac{1}{2}}} \la(x)~,   
\ee   
\be\mlab{lp51c}   
\rho_{in}(x)  \rar   \rho_{in}(x)~,~~  ~ 
U^\mu_{in}(x) \rar  U^\mu_{in}(x)  ~.  
\ee   
The global transformation $\phi_H(x) \rar e^{i  \te} \phi_H(x)$
is induced by    
\be\mlab{lp53a}   
\chi_{in}(x)  \rar  \chi_{in}(x) \, + \, \frac{{\ti   
v}}{Z_\chi^{\frac{1}{2}}} \te  ~,
\ee
\be\mlab{lp53b}   
b_{in}(x)  \rar   b_{in}(x) ~, ~  
\rho_{in}(x)  \rar   \rho_{in}(x) ~,~  
U^\mu_{in}(x)  \rar   U^\mu_{in}(x) ~. 
\ee   
Note that under the above in-field transformations  
the in-field equations and the $S$    
matrix  are invariant and that $B_H$ is changed    
by an irrelevant c-number. We thus see that the original 
invariance cannot be lost even at the level of the    
physical fields, although it can manifest there    
in a different symmetry group structure. 

Eq. (\ref{lp53a}) shows that the physical field $\chi_{in}$ translates    
by a constant when the Heisenberg field    
$\phi$ undergoes the global phase transformation (and vice versa):  
The global $U(1)$ invariance group is dynamically rearranged into    
the one-parameter constant translation group. This last one is    
the group contraction of global $U(1)$.   
   
Notice that translation    
by a constant is an invariant transformation for   
the $\chi_{in}$ field equation if and only if $\chi_{in}$ is a    
massless field. Thus the rearrangement into the contraction of    
global $U(1)$ group has the same content as the Goldstone 
theorem\cite{dv}. 
It is also interesting to note that while $U(1)$ is a compact    
group its contraction is not compact. Since   
the number operator of $\chi_{in}$ field changes under the    
translation (\ref{lp53a}) we say that we have $\chi_{in}$ coherent   
boson condensation.   
  
It must be stressed that the translation by a constant   
(\ref{lp53a}) must be actually understood as the limit for   
$f(x) \rar 1$ of the transformation    
\be\mlab{gp36a}   
 \chi_{in}(x,\te) \, =\, \chi_{in}(x) \, + \,   
\frac{\ti v}{Z_\chi^{\frac{1}{2}}} \te f(x) ~,   
\ee   
with $f(x)$ a    
normalizable solution of the $\chi_{in}$ field equation:    
$\pa^{2} f(x)=0$. Eq. (\ref{lp53a})
induces infrared    
singularities in Feynman diagrams with many momentumless and    
energyless lines. These are smeared out by use of (\ref{gp36a}). 
However, notice that matrix    
elements are well defined even when (\ref{lp53a}) is used. 
Also, the function $f(x)$    
appearing in the generator of (\ref{gp36a}) makes it 
well defined\cite{MPU742,sha,wy}.    
   
Since different physical phases (unitarily inequivalent representations)   
are associated to different NG boson   
condensation densities, we see that boson translations, by inducing   
variation of NG boson condensation, 
represent transitions through physically different    
phases. In particular, I will discuss non-homogeneous boson condensation
induced by transformations such as (\ref{gp36a}).

The dynamical rearrangement of symmetry has been studied in many    
models of physical interest. In the case of global $SU(2)$    
invariance group, for example in ferromagnets\cite{sha}, 
or in systems with    
isospin vector fields\cite{wy}, the NG boson condensation is controlled by    
the $E(2)$ group, which is the group contraction of $SU(2)$\cite{dv}.    
Unfortunately, for lack of space I cannot report about these and    
some other interesting cases.

\section{Observable effects of non-homogeneous boson transformations}

Translations    
of bosonic physical fields (not necessarily massless) by    
space-time dependent functions, say $\al (x)$, 
satisfying the    
same field equation of the translated physical field, are called    
{\it boson transformations}\cite{Um2,Um1,MPU742}. Eq. 
(\ref{gp36a}) is thus an example of boson transformation.

Let $\phi'_H$ denote the Heisenberg   
field obtained through the dynamical mapping when the physical   
field undergoes the boson transformation. The {\it boson transformation    
theorem} can be then proved, which states that $\phi'_H$ is also   
a solution of the Heisenberg field equation\cite{Um2,Um1,MPU742}.   
   
The proof of the theorem consists in showing that   
the boson-transformed fields, say
$\phi_H[x;\chi_{in}(x)  + \al(x)]$,
differ from $\phi_H[\chi_{in}(x)]$  
only by an $\ep-$dependent factor and therefore are    
solutions of the same field equations.   
   
In the absence of a gauge field, under boson transformation the order 
parameter gets space-time dependence given by   
\be\mlab{gp27}   
{\ti v}(x)      
= \exp{\le\{i\frac{Z_{\chi}^{\frac{1}{2}}}{\ti v} \al(x)\ri\}}   
\le[{\ti v} + V \le(i\frac{Z_{\chi}^{\frac{1}{2}}}{\ti v}    
\pa_\mu\al(x)\ri)\ri]  ~, 
\ee   
where the expansion of ${\ti v}(x)$ around $\pa_\mu\al(x)=0$ is used   
($V \le(i\pa_\mu\la(x)\ri) \rar 0$ when   
$\pa_\mu\al(x)\rar 0$). Note that the modulus of $\ti v$   
changes and in the limit $\al \rar   
const.$ only its phase is changed.  
  
In the case a gauge field is present, it can be shown that 
any space-time 
dependence of the $\ep-$term can be eliminated by a gauge   
transformation when $\al (x)$ is a regular (i.e. Fourier   
transformable) function and the only effect is the appearance  
of a phase factor in the order parameter:  
${\ti v}(x)\, =\, e^{ic\al (x)} {\ti v}$, with $c$ a constant.  
  
The conclusion is that when a gauge field is present, the boson    
transformation   
with regular $\al (x)$ is equivalent to a   
gauge transformation. On the contrary, in a theory with global    
invariance only, non-singular boson   
transformations of the NG fields can produce 
non-trivial physical effects (like linear flow in superfluidity).   
   
I want to stress that, in the case of global phase    
transformations as    
well as in the case of local gauge transformations, the proof of    
the boson transformation theorem relies on the fact that $\al (x)$    
is a regular function. If one wants to consider functions with    
some singularities (divergence and topological singularities) one    
has to carefully exclude the singularity regions when integrating on    
space and/or time. For example, if $\al (x)$ is singular on the axis    
of a cylinder (at $r =0$) one must exclude the singular line $r =    
0$ by a cylindrical surface of infinitesimal radius. The phase of    
the order parameter will be singular on that line. This means    
that SBS does not occur in that region (the core): there we have    
the "normal" state rather than the ordered one.   
Provided one uses such care, the boson transformation can    
be safely (and advantageously) used also in the case of singular $\al(x)$.   
   
The  
boson theorem has relevant physical meaning    
since it shows that the same dynamics may describe homogeneous    
and non-homogeneous phenomena.
When a theory allows SBS, there    
always exist solutions of the field   
equations with space and/or time-dependent vacuum. These solutions are   
obtained from the translationally invariant ones by the boson   
transformation of the NG field: they results from a local Bose   
condensation of the $\chi_{in}$  particles.   
This directly leads us to the    
mechanism of formation of extended objects (defect formation).   

Notice that in local gauge theories the boson    
transformation must be compatible     
with Heisenberg field    
equations but also with the physical state condition (\ref{lp49}).   
Under the boson transformation with $\al (x) =  \frac{{\ti   
v}}{Z_\chi^{\frac{1}{2}}} f(x)$ and 
$\pa^2 f(x) =0$, $B_H$ changes as   
\be\mlab{vs9}   
B_H(x) \rar B_H(x) - \frac{e_0 {\ti v}^2}{Z_\chi} f(x) ~,   
\ee  
Eq. (\ref{lp45}) is thus violated when the
Gupta-Bleuler-like condition is imposed. In order to    
restore it, the shift in $B_H$ must be compensated by means of   
the transformation on $U_{in}$:   
\be\mlab{vs10}   
U^{\mu}_{in}(x) \rar U^{\mu}_{in}(x) +   
{Z_{3}}^{-\frac{1}{2}} a^{\mu}(x) \qquad ,   
\qquad \pa_\mu a^{\mu}(x)=0 ~,  
\ee   
with a convenient c-number function 
$a^{\mu}(x)$.    
The dynamical   
maps of the various Heisenberg operators are not    
affected by (\ref{vs10}) since they contain   
$U^{\mu}_{in}$ and $B_H (x)$ in a combination such that the changes of    
$B_H$ and of $U^{\mu}_{in}$ compensate each other provided   
\be\mlab{vs20}   
(\pa^2 + m_V^2) a_\mu(x) \, = \,\frac{m_V^2}{ e_0} \pa_\mu f(x)~.   
\ee   
Eq. (\ref{vs20}) thus obtained is the Maxwell equation for   
the massive potential vector $a_{\mu}$\cite{MPUV75,MPU75}. 
The classical ground   
state current $j_{\mu}$ turns out to be   
\be\mlab{vs21}   
j_\mu(x)\equiv \lan 0| j_{H \mu}(x) |0 \ran \, =\,   
 m_V^2 \le[ a_\mu(x) - \frac{1}{e_0} \pa_\mu f(x) \ri]~.   
\ee   
The term $ m_V^2  a_\mu(x)$ is the {\em Meissner current}, while    
$ \frac{m_V^2}{e_0} \pa_\mu f(x)$ is the {\em boson current}.    
   
{\it The macroscopic field and current are thus given in terms of the    
boson condensation function}.  
  
The (classical) boson current is given by 
${\pa}_{\mu}f$, i.e. by variations in the    
non-homogeneous boson condensate: boson condensation functions    
must play a r\^ole in phase transitions    
where boson condensate indeed changes.  
  
{\it Let me now show that boson transformation functions carrying 
topological   
singularities are only allowed for massless bosons}\cite{Um1,MSU79,wa}.  
  
Suppose the function $f(x)$ for the boson transformation of the   
field $\chi_{in}$ carries a topological singularity and is  
thus path-dependent:   
\be   
\\ \mlab{ts1}   
G^{\dag}_{\mu\nu}(x) \equiv [\pa_\mu,\pa_\nu]\,f(x) \neq 0~,  
\qquad for \;certain \; \mu\, , \,  \nu   
\, , \, x  ~. 
\ee  
On the other hand, $\pa_\mu \, f$, which is related with   
observables, is single-valued,  
i.e. $[\pa_\rho,\pa_\nu]\,\pa_\mu f(x)\,=\,0$. Recall that  
$f(x)$ is solution of the $\chi_{in}$ equation and suppose there is   
a non-zero mass term:   
\be\mlab{ts3}  
(\pa^2 + m^2)f(x) = 0~.  
\ee  
From the definition of $G^{\dag}_{\mu\nu}$ and   
the regularity of $\pa_\mu f(x)$ it follows that  
\be\mlab{ts5}   
\pa_\mu f(x) \, =\, \frac{1}{\pa^2 \, + \, m^2}   
\pa^{\la} \,G^{\dag}_{\la\mu}(x) ~,  
\ee   
which leads to   
$\pa^2 f(x) \, =\,0$ which implies $m=0$.   
   
Thus (\ref{ts1}) is only compatible with   
massless equation for $\chi_{in}$.  
   
The quantity $\pa_\la \,f(x)$ is given by (\ref{ts5}) with $m=0$.  
From this equation, $f(x)$ can be determined.     
The topological charge is defined as   
\be\mlab{ts8}  
N_T=\int_C \,dl^\mu\, \pa_\mu \, f   
= \int_S\, dS_{\mu}\ep^{\mu\nu\si}\, \pa_\nu \pa_\si \, f\, =\,   
\frac{1}{2}\int_S\, dS^{\mu\nu}\, G^{\dag}_{\mu\nu} ~.  
\ee   
Here $C$ is a contour enclosing the singularity and $S$ a surface with   
$C$ as boundary. $N_T$ does not depend  on the path $C$ provided this   
does not cross the singularity. The dual tensor $G^{\mu\nu}$ is    
\be\mlab{ts9}   
G^{\mu\nu}(x)\equiv - \frac{1}{2}\,   
\ep^{\mu\nu\la\rho}G^{\dag}_{\la\rho}(x)    
\ee   
and satisfies the continuity equation   
\be\mlab{ts10}   
\pa_\mu\,G^{\mu\nu}(x)\,=\,0    
\qquad \Lrar \qquad    
\pa_\mu\,G^{\dag}_{\la\rho}\, +\,\pa_\rho\,G^{\dag}_{\mu\la}\, +\,   
\pa_\la\,G^{\dag}_{\rho\mu}\,=\,0   ~.
\ee   
This equation completely characterizes the topological singularity   
\cite{Um1,wa}.    
  
Let me now observe that all the macroscopic ground state effects   
do not occur for regular $f(x)$ ($G^{\dag}_{\mu\nu} = 0$). In   
fact, from (\ref{vs20}) we obtain $a_{\mu}(x) = \frac{1}{e_{0}}  
\pa_{\mu} f(x)$ for regular $f$ which implies zero classical   
current ($j_{\mu} = 0$) and zero classical field   
($F_{\mu\nu} = \pa_{\mu} a_{\nu} -  \pa_{\nu} a_{\mu}$), since the   
Meissner and the boson current cancel each other.  
  
In conclusion, the vacuum current appears only when $f(x)$ has   
topological singularities and these can be created   
only by condensation of massless bosons, e.g. by condensation of NG
bosons when SBS occurs.  
This explains why topological defects appear in the process of   
phase transitions, where NG modes are present and gradients in   
their condensate densities are nonzero.  
  
On the other hand, the appearance of space-time order parameter   
is no guarantee that persistent ground state currents (and   
fields) will exist: if $f$ is a regular function, the space-time   
dependence of $\ti v$ can be gauged away by an appropriate gauge   
transformation.   
  
Since the boson transformation with   
regular $f$ does not affect observable quantities, the $S$  
matrix (\ref{lp56a}) is actually given by  
\be\mlab{lp56aa}   
S\,=\,: S[\rho_{in}, U^\mu_{in} - \frac{1}{m_V} \pa(\chi_{in} -  
 b_{in})] : ~.    
\ee   
This is in fact independent of the boson transformation with   
regular $f$:  
\be\mlab{lp56ab}   
S\,\rar\,S' = : S[\rho_{in}, U^\mu_{in} - \frac{1}{m_V}   
\pa(\chi_{in} - b_{in})  
+ Z^{-\frac{1}{2}}_{3} (a^{\mu} - \frac{1}{e_{0}} {\pa}^{\mu} f)]:    
\ee   
since  $a_{\mu}(x) = \frac{1}{e_{0}}  
\pa_{\mu} f(x)$ for regular $f$. However, $S'   
\neq S$ for singular $f$: $S'$ includes the interaction of the quanta   
$U^\mu_{in}$ and $\phi_{in}$ with the defect classical field   
and current.   
  
Thus we see how quantum fluctuations may interact and have effects on   
classically behaving macroscopic defects:    
our picture includes interaction of quanta   
with macroscopic objects.  
Much more can be said on the interaction of extended objects with   
quanta; however, for shortness I will not discuss more on   
that.  
 
\section{The vortex solution}  
   
The meaning of Eq. (\ref{ts5}) with $m = 0$ is of 
course\cite{Um1,MPU75,wa,tze}  
\be\mlab{vs22}   
\pa_{\mu} f(x) \, =\, 2 \pi \int d^4 x' \, G^{\dag}_{\mu \nu}(x')   
\pa_x^\nu K(x-x')~,    
\ee   
\be\mlab{vs23}   
K(x-x')\, =\, - \frac{1}{(2 \pi)^4} \int d^4 p \,    
e^{-i p (x-x')} \frac{1}{p^2 + i\ep} ~,   
\ee   
with the Green's function satisfying $\pa^2 K(x-x')   
=\de(x-x')$. Eq.(\ref{vs22}) gives upon path integration  
\be\mlab{vs24}   
f(x) \, =\, 2 \pi \int^x dx^\mu \int d^4 x' \, G^{\dag}_{\mu \nu}(x')   
\pa_x^\nu K(x-x') ~,    
\ee   
which is indeed solution of $\pa^{2}f(x) = 0$.  
The classical vector potential is 
\be\mlab{vs25}   
a_\mu(x) \, =\, -\frac{m_V^2}{e}\int d^4 x' \,    
\De_c(x-x') \pa_\mu'f(x')~,    
\ee   
\be\mlab{vs26}   
\De_c(x-x')\, =\, \frac{1}{(2 \pi)^4} \int d^4 p \,    
e^{-i p (x-x')} \frac{1}{p^2 - m_V^2 + i\ep} ~.  
\ee   
The electromagnetic tensor and the vacuum current are\cite{Um1,MPU75,wa}   
\be\mlab{vs27}   
 F_{\mu \nu}(x) = \pa_\mu a_{\nu}(x) - \pa_\nu   
a_{\mu}(x)= 2\pi \frac{m_V^2}{e}\int d^4 x' \,    
\De_c(x-x') G^{\dag}_{\mu \nu}(x')  ~,  
\ee   
\be\mlab{vs28}   
j_\mu(x) \, =
=\, - 2\pi \frac{m_V^2}{e}\int d^4 x' \,     
\De_c(x-x') \pa_{x'}^\nu G^{\dag}_{\nu \mu}(x') ~,  
\ee   
respectively, and satisfy $\pa^\mu F_{\mu \nu}(x)\, =\,-j_\nu(x)$.  
  
The line singularity for the vortex solution can be parameterized   
by a single line parameter $\si$ and by the time parameter   
$\tau$. The {\it static vortex} solution is   
then obtained by setting $y_0(\tau,\si)=\tau$   
and ${\bf y}(\tau,\si)={\bf y}(\si)$, with $y$ denoting the line   
coordinate. $G^{\dag}_{\mu \nu}$ is non-zero only on the line   
at $y$ (we can consider more lines but here I limit myself to one   
line, for simplicity). Thus,  
\be\mlab{vs31a}   
G_{0i}(x) \, =\,\int\, d\si\,   
 \frac{d y_i(\si)}{d \si}   
 \, \de^{3}[{\bf x} - {\bf y}(\si)] ~,~ ~~
G_{ij}(x) \, =\,0~,  
\ee
\be\mlab{vs31b}   
G_{ij}^{\dag}(x) \, =\,-\ep_{ijk}G_{0k}(x)   
\qquad, \qquad   
G_{0i}^{\dag}(x) \, =\,0 ~.  
\ee 
Eq.(\ref{vs27}) shows that these vortices are purely {\em magnetic}.  
Eq.(\ref{vs22}) gives    
\be\mlab{vs32b}   
\pa_{i}f(x) \, =\,\frac{1}{(2\pi)^2}\int\, d\si\,   
\ep_{ijk} \frac{d y_k(\si)}{d \si} \pa_j^x \int d^3p   
\frac{e^{i {\bf p}\cdot({\bf x}-{\bf y}(\si))}}{{\bf p}^2} ~,  
\ee  
and $\pa_{0}f(x) \, =\, 0$, i.e., by using the identity $(2\pi)^{-2}\int   
d^3p \frac{e^{i {\bf p}\cdot{\bf x}}}{{\bf p}^2}= \frac{1}{2 |{\bf x}|}$,   
\be   
\mlab{vs33}   
\nabla f(x) \, =\,-\frac{1}{2}\int\, d\si\,   
\frac{d {\bf y}_k(\si)}{d \si} \wedge \nabla_x   
\frac{1}{|{\bf x}-{\bf y}(\si)|} ~. 
\ee   
Note that $\nabla^2 f(x) =0$ is satisfied.  
  
{\it Straight infinitely long vortex} is specified by   
$y_i(\si)= \si\, \de_{i3}$ with $ -\infty < \si < \infty$.  
The only non vanishing component of $G^{\mu\nu}(x)$ are  
$G^{03}(x)=G_{12}^{\dag}(x)\,=\, \de(x_1)\de(x_2)$.  
Eq.(\ref{vs33}) gives\cite{Um1,MPU75,wa}
\be\mlab{vs36a}   
\frac{\pa}{\pa x_1} f(x) =\frac{1}{2}\int d\si \frac{\pa}{\pa x_2}    
[x_1^2 + x_2^2 + (x_3 -\si)^2]^{-\frac{1}{2}}=   
- \frac{x_2}{x_1^2 + x_2^2} ~,  
\ee
\be\mlab{vs36b}   
 \frac{\pa}{\pa x_2} f(x) =\frac{x_1}{x_1^2 + x_2^2} ~,~~~  
\frac{\pa}{\pa x_3} f(x)  = 0 ~,  
\ee   
which give   
\be   
\mlab{vs37}   
f(x)\, =\, \tan^{-1} (\frac{x_2}{x_1})=\te(x)   ~.
\ee   
Use of these results gives the vector potential, $F_{\mu \nu}$   
and the vacuum current. The only non-zero components of these   
fields are $a_{1}$, $a_{2}$, $F_{1 2}$, $j_{1}$ and   
$j_{2}$.   
  
Notice that    
the condition (\ref{ts10}) can be shown to be violated if the line   
singularity has isolated end   
points inside the system. Thus consistency with the   
continuity equation (\ref{ts10}) implies that either the string is   
infinite, or that it form a closed loop. Also, if there are more than   
one string, the end points of different strings can be connected in a   
vertex, eq.(\ref{ts10}) resulting then in a condition for the relative   
string tensions $\nu_\al$, with $\al$ denoting different strings.   
   
Further simple examples are the following\cite{Um1,MPU75,wa}.   
  
A {\it circular loop}:   
$\vec{y}(\si)= (a \cos\si, a \sin\si,0)$, $0 \leq \si \leq   
2\pi$.  
$G^{01}(x) = \de\le[x_2-\sqrt{a^2-x_1^2}\ri]    
 \,\de(x_3)$, ~ $G^{02}(x) = \de\le[x_1-\sqrt{a^2-x_2^2}\ri]    
 \,\de(x_3)$, ~ $G^{03}(x) = G^{ij}(x)\, =\, 0$.   
   
A straight line along the third axis moving in the $x_1$ direction with   
velocity $v$ is given by $\vec{y}(\si,\tau)= (v\tau,0,\si)$,   
$y_0(\si,\tau) =\tau$, from which $G^{03}(x) = \nu\,\de(x_1-vt)    
 \,\de(x_2)$ and $G^{13}(x) = v\,\de(x_1-vt) \,\de(x_2)$.

\section{Finite temperature and finite volume effects}   

Consider the $U(1)$ invariant model at finite temperature.
The breakdown of symmetry condition in the  
homogeneous condensation case is\cite{MV90}  
\be\mlab{v.1}  
\lan0(\beta)|\phi(x)|0(\beta)\ran = {1\over {\sqrt{2}}} \si(\beta)~,  
\ee  
where $\beta \equiv \frac{1}{K_{B}T}$. $|0(\beta)\ran$ denotes the  
temperature dependent vacuum state in Thermo Field Dynamics  
\cite{Um1}. Note that the statistical average of any operator $A$  
is given by $<A>_{0} = \lan0(\beta)|A|0(\beta)\ran$.

I omit to consider here the presence of other fields (such as   
the ghost fields) for brevity.  
The fields $\phi$, $\chi$ and $A_{\mu}$ may undergo translation   
transformations by c-number functions, say $\sigma$, $\kappa$ and  
$\alpha_{\mu}$, respectively, controlling the respective   
condensate structures. I write $\phi \equiv \rho + {1\over   
{\sqrt{2}}} \si(\beta)$. Usual   
gauge transformations are induced by using $\sigma = 0$, $\kappa   
= \alpha (x)$ and $\alpha_{\mu} (x) = {\pa}_{\mu}\alpha (x)$.  
  
The homogeneous boson condensation of the Higgs field alone   
($\sigma (\beta) = {\sigma}_{0}(\beta) = const. \not= 0$, $\kappa = 0$   
and $\alpha_{\mu} = 0$) leads to   
\be\mlab{v.2}  
m^{2} = 2\la \si^{2}_{0} ~~, ~~~ 
M^{2} = e^{2}({\si^{2}}_{0} + <:\bar{\rho}^{2}:>_{0}) ~~,
\ee
\be\mlab{v.2x}
\si^{2}_{0} = v^{2} -3<:\bar\rho^{2}:>_{0} + {e^{2} \over   
\la}<:{\bar A}_{\mu}{\bar A}^{\mu}:>_{0} ~~,  
\ee  
where $m$ and $\la$ denote the Higgs field mass and self-coupling,   
respectively, $v \equiv \lan 0|\phi|0 \ran$ at $T = 0$ and is   
assumed to be non-zero, $M$ is   
the gauge field mass and $e$ is the (electric charge) coupling   
between $A_{\mu}$ and $\phi$. ${\bar\rho}$ and ${\bar A}$ denote the
physical fields.
  
Eqs.(\ref{v.2}) are actually   
self-consistent equations since $<:\bar\rho^{2}:>_{0}$  
also depends on $m^{2}$. In the discontinuous phase transition   
case the free energy should be   
examined\cite{MV90}. The proper phase transition point is defined by the   
equality between the ordered and the disordered free energy phase.  
  
As $T \rar 0$ eqs. (\ref{v.2}) show that $\si_{0} \rar v$ thus   
recovering the original zero temperature symmetry breaking. 
We have phase transition to the (disordered)   
phase $\si_{0}(\beta_{C}) = 0$ at the critical temperature $T_{C}$  
such that  
\be\mlab{v.3}  
v^{2} = 3<:\bar\rho^{2}:>_{0} - {e^{2} \over   
\la}<:{\bar A}_{\mu}{\bar A}^{\mu}:>_{0} ~~.  
\ee  

Above the phase transition point $T > T_{C}$, and   
$\si_{0} = 0$,   
we have\cite{MV90}  
\be\mlab{v.4}  
m^{2} = 
-\la v^{2} + 3\la <:\bar\rho^{2}:>_{0} - e^{2}   
<:{\bar A}_{\mu}{\bar A}^{\mu}:>_{0} ~~.  
\ee  
Full symmetry restoration (i.e. $v = 0$) occurs at   
$T \equiv T_{sym}$ such that thermal contributions in (\ref{v.3})  
compensate each other, and then also $m = 0$.  
The gauge field mass $M$ goes to   
zero not at $T_{C}$, but at $T$ such that   
\be\mlab{v.5}  
\la v^{2} - 2\la <:\bar\rho^{2}:>_{0} + e^{2}   
<:{\bar A}_{\mu}{\bar A}^{\mu}:>_{0} = 0 ~~.  
\ee  

The vortex solution arises in the non-homogeneous condensation  
case obtained by introducing space dependence in the   
condensate functions. 
Introducing the cylindrical coordinates, the asymptotic gauge   
field configuration is imposed by considering the    
angle function as gauge function at infinity  
$k(x)\, =\, \frac{n}{e}\, \te$, ~ ($k(x) = 0$ at $r=0$): 
\be\mlab{ve3}  
\al^i_{as}=-\frac{n}{er}\, {\bf e}^i_{\te} ~. 
\ee  
Here $n$ is the winding number and we see that,   
although, as already observed,   
the NG bosons do not enter the physical spectrum,   
nevertheless their condensation is directly related to the   
topological charge.  
For $r<\infty$ we assume (the vortex ansatz)\cite{NO}   
\be\mlab{ve4}   
\al^i\, =\, -\frac{n}{er}(1-K(r))\,{\bf e}^i_{\te}~,~~~ 
\si(x)\, =\, \si_0f(x) ~,  
\ee 
where $\si_0$ is the Higgs field shift for the homogeneous condensation   
and    
\be\mlab{ve5}   
K(r)=1-\frac{r}{n}A(r), \qquad A(r)\rar \frac{n}{r}\,\,\,~ for\,\,\,
~ r\rar\infty   ~.
\ee   
The vortex ansatz leads to the  
temperature dependent vortex equations   
\bea\mlab{ve6}   
&&  \frac{d}{dr}\le(\frac{1}{r}\,    
\frac{d}{dr}(rA)\ri)=e^2\le(A-\frac{n}{r}\ri)(\lan:\bar{\rho}^2:   
\ran_0+f^2\si^2_0) ~ , \\     
&& \frac{1}{r}\, \frac{d}{dr}\le(   
r\frac{df}{dr}\ri)=\le(A-\frac{n}{r}\ri)^2f-\la\si_0^2f(1-f^2) ~.  
\eea   
As $T\rar 0$ these equations reduce to the usual vortex equations.   
One can show\cite{MV90} that in the vortex case the masses are given by  
\be\mlab{v.7}  
m^{2}(x) = 2\la \si^{2}_{0}f^{2}(x) ~~,\\ ~~~ 
M^{2}(x) = e^{2}(\si^{2}_{0}f^{2}(x) + <:\bar\rho^{2}:>_{0}) ~~.  
\ee  
These masses act as potential terms in the field equations and   
only at spatial infinity ($r \rar \infty , f(x) \rar 1$) ordinary   
mass interpretation is recovered. We have in fact the asymptotic   
behavior  
\be\mlab{v.8}  
K(r)\simeq e^{-Mr} = e^{- {r \over{R_{0}}}} ~~,\\ ~~~ 
f(r)\simeq 1 - f_{0}e^{-mr} = 1- f_{0}e^{- {r \over {r_{0}}}}~~.  
\ee  
$R_{0}\equiv {1 \over M}$ gives the size of the gauge field   
core and $r_{0} \equiv {1 \over m}$ the Higgs field core.  
As $T   
\rar T_C$ the Higgs field core increases and the gauge field core   
becomes smaller. At $T=T_C$ one obtains the pure gauge field   
core. Above $T_C$ symmetry is restored. The discussion on   
temperature dependence of $\si_{0}$ is similar to the one for the   
homogeneous case.  
  
The 't Hooft-Polyakov monopole and the sphaleron   
solutions at finite temperature are discussed in \cite{MV90}. 
  
Let me now discuss the effects of the finite size of the system  
on the boson condensate and the relation between finite size and   
temperature effects. This will help to understand how  
temperature variations near $T_C$ control the defect size (and  
thus the defect number)\cite{ro}. 
  
In the case of large but finite volume we expect that the  
condition of symmetry breakdown 
is still satisfied ``inside the bulk'' {\em far} from the  
boundaries. However, ``{\em near}'' the boundaries, one might  
expect ``distortions'' in the order parameter: ${\ti v} ={\ti  
v}(x)$ (or even ${\ti v}\rar 0$). ``Near''  
the system boundaries we may have non-homogeneous order parameter.  
Non-homogeneities in the boson condensation   
will ``smooth out'' in the $V\rar\infty$ limit.  
Suppose the integration in eq. (\ref{gp6}) is over the 
finite (but large) volume $V \equiv \eta ^{-3}$ and use  
\be\mlab{note13}  
\de_\eta(p)=\frac{1}{2\pi}\int_{-\frac{1}{\eta}}^{\frac{1}{\eta}}  
dx\,e^{ipx}=\frac{1}{\pi p}\, sin\frac{p}{\eta}  ~, 
\ee  
which, as well known, approaches $\de (p)$ as $\eta\rar 0$: $  
lim_{\eta\rar 0} \de_\eta(p)=\de (p) $. Now  
\be\mlab{note15}  
lim_{\eta\rar 0} \int dp\, \de_\eta(p)\,f(p)=f(0)=   
lim_{\eta\rar 0} \int dp\, \de(p-\eta)\,f(p)  ~,
\ee  
then, using  
$\de_\eta(p)\simeq\de(p-\eta)$ for small $\eta$, it is   
\be\mlab{6}  
{\ti v}(\vec{x},\ep , \eta)=i\ep  
e^{-i\vec{\eta}\cdot\vec{x}}\, \Delta_{\chi}(\ep,\vec{\eta},p_0=0)  ~,
\ee  
\be\mlab{gp8}   
\De_\chi(\ep, \vec{\eta}, p_{0}=0) =  
\lim_{\ep \rar 0}\le[ \frac{Z_\chi}{{- \omega_{\vec{p} =  
\vec{\eta}}^{2}} 
+i\ep  a_\chi} + c.c. \ri]\, .   
\ee   
with ${- \omega^{2}}_{\vec{p} = \vec{\eta}} = \vec{\eta}^{2} +  
{m^{2}}_{\chi}$. Thus, ${lim_{\ep \rar 0}} {lim_{\eta \rar  
0}} {\ti v}(\vec{x}, \ep, \eta) \neq 0$ only if $m_{\chi} = 0$,  
otherwise ${\ti v} = 0$. Eq. (\ref{gp9}) (the Goldstone theorem) 
is thus recovered in the infinite volume limit ($\eta \rar 0$).  
 
On the other hand, assume that NG modes are there,  
i.e. $m_{\chi} = 0$, then $\omega_{\eta} \equiv  
{- \omega}_{\vec{p} = \vec{\eta}} \neq 0$ for $\vec{\eta} \neq 0$  
and it acts as an "effective mass" for the NG boson. Here I mean  
that the NG boson is massless, but its lowest (zero) energy is  
prevented to be reached, the lowest energy being given by the  
non-zero $\omega_{\eta}$. The effect of the  
boundaries ($\eta \neq 0$) is to give an "effective mass" (in the  
above sense) $m_{eff} \equiv\omega_{\eta}$ to the NG bosons.  
These will then propagate over a range of the order of  
$\xi\equiv\frac{1}{\eta}$, which is the system linear size. 
  
Notice that only if  
$\ep \neq 0$ the order parameter can be kept different from zero, 
i.e. if  $\eta \neq 0$ then $\ep$ must be  
non-zero in order to have ${\ti v} \neq 0$ (at least locally). In  
such a case the symmetry breakdown is maintained thanks to the 
non-zero coupling, $\ep \neq 0$, with an external field acting  
as an external pump providing  
energy: energy supply is required in order to condensate   
modes of non-zero lowest energy $\omega_\eta$.  
Boundary effects are thus in competition with  
breakdown of symmetry\cite{ro,ko}. They may preclude its occurrence or, if  
symmetry is already broken, they may reduce to zero the order  
parameter. 
 
We have seen that temperature may have similar effects on the  
order parameter (at $T_{C}$ symmetry may be restored, cf. eq.  
(\ref{v.3}) and the discussion following it). Since the order  
parameter goes to zero when NG modes acquire non-zero effective  
mass, we may represent the effect of thermalization in terms of  
finite volume effects and put, e.g., $\eta \propto  
{\sqrt{\frac{|T-T_C|}{T_C}}}$. In this way temperature  
fluctuations around $T_{C}$ may be discussed as fluctuations in the 
condensate domain size $\xi$. For example, in the presence of an  
external driving field ($\ep \neq 0$), for $T > T_{C}$ but near to  
$T_{C}$ one may have the formation of ordered domains of size  
$\xi \propto ({\sqrt{\frac{T-T_C}{T_C}}})^{- 1}$  {\it before}  
phase transition to fully ordered phase is  
achieved (as $T \rar T_{C}$). As far as $\eta \neq 0$,   
the ordered domains are unstable, they disappear as the   
external field coupling $\ep \rar 0$.    
  
Of course, if ordered domains are still   
presents at $T<T_C$,  
they also disappear as $\ep\rar 0$.  
The possibility to maintain such ordered   
domains below $T_C$ depends on the speed by which $T$  
is lowered, compared to the speed by which the system is able to  
get homogeneously ordered. Notice that the speed by which   
$T \rar T_C$ is related to the speed by which $\eta \rar 0$.  
 
In the case of the kink solution it can be shown\cite{MV1} that 
the mass $\mu_0=(2\la )^{\frac{1}{2}}v(\beta)$   
of the "constituent" fields $\rho^{in}$ fixes the kink   
size $\xi_\beta\propto\frac{2}{\mu_0}=\frac{\sqrt{2}}{\sqrt{\la}v(\beta)}$  
which thus increases as $T\rar T_C$. It is also interesting to note that   
in the $T \rar 0$ limit the kink size is  
$\xi_0\propto\frac{\sqrt{2}}{\sqrt{\la}v}<  
\frac{\sqrt{2}}{\sqrt{\la}v(\beta)}=\xi_\beta$,  
since $v^2(\beta) <v^2$.   

As $T$ is different from zero, the  
thermal Bose condensate $\lan:\rho^2:\ran_0$ develops which acts as  
a potential term for the  
quantum field   
$\rho(x)$. Such   
temperature effects manifest at classical level as potential term   
for the classical kink field.  
It is such a potential term which actually controls   
the ``size'' (and the number) of the  
kinks. Notice that  $\mu_0^2(x,\beta)$ also acts as a  
potential for   
the $\rho^{in}_\beta(x)$ field. Only in the limit $v(x,\beta)\rar const$  
the $\rho^{in}_\beta(x)$ field may be considered as a free field;  
e.g. far from the kink core.   
  
The  $\rho^{in}_\beta(x)$ condensation,   
whose macroscopic envelope is represented by the soliton   
solution, is induced by the boson transformation with  
$f_{\beta}(x)=const.\, e^{-\mu_0(\beta)x_1}$ playing the role of ``form  
factor''. The number of condensed bosons is thus proportional to   
$\vert f_{\beta}(x)\vert^2 = e^{-2\mu_0(\beta)\,(x_1-a)}$,
which is maximal near the kink center $x_1=a$ and decrease over a size  
$\xi_\beta=\frac{2}{\mu_0(\beta)}$. The meaning of the boson  
transformation is that the $f_{\beta}$-translation breaks   
the homogeneity of the otherwise constant in space order   
parameter $v(\beta)$.

In conclusion, phase transitions imply "moving" over unitarily   
inequivalent representations, and this in general implies  
non-trivial homotopy mapping between the $(x,\beta)$ variability   
domain and the group manifold. 
The order parameters $v(x, \beta)$ and $\sigma (x, \beta)$  
introduced above provide a mapping between the variation domains  
of $(x, \beta)$ and the {\it space of the unitarily inequivalent  
representations} of the canonical  
commutation relations, i.e. the set of Hilbert spaces where the  
operator field $\phi$ is realized for different values of the   
order parameter.
The invariance under   
the theory symmetry group then {\it necessarily} leads to NG    
boson condensation functions with topological singularities. 
In other words, since phase transitions imply observable changes in the
system physical properties, boson condensation functions relevant to
phase transitions necessarily carry topological singularities
(non-singular functions have no observable effects, indeed). This
explains why we observe defect formation in the process of phase transition.
 
In the case of the kink 
there are no NG modes, nevertheless the   
topologically non-trivial kink solution requires the boson   
condensation function to carry divergence   
singularity (at spatial infinity).  
 
This work has been partially supported by   
INFM, MURST and the ESF Network on Topological defect formation in  
phase transitions.

\end{document}